# Impedance-based AC/DC Terminal Modeling and Analysis of the MMC-BTB System

Chongbin Zhao, *Student Member, IEEE*, and Qirong Jiang

*Abstract*—Impedance-based small-signal stability analysis is widely applied in practical engineering with modular multilevel converters (MMCs). However, the deficiencies of existing impedance models (IMs) and the idealized extension for the single MMC influence the analyses in multiterminal systems. Such gaps are filled by focusing on an MMC-based back-to-back system in this paper. To obtain the steady-state trajectory of the system, a numerical method is first proposed based on Newton–Raphson iteration in the frequency domain. Then, by substituting the shared terminal dynamics with active or passive devices, theoretical AC/DC IMs that consider typical control loops with the pure time delay, are directly established based on the multiharmonic linearization. Further aided by the derived IMs, two neglected aspects in the current literature, i.e., the influence of power transformers on low-frequency impedance characteristics and the rationality of using simplified IMs for high-frequency resonance studies, are investigated. It is confirmed that the stability of interlinking systems should be comprehensively analyzed at both AC and DC terminals. This helps position the instability source, obtain the stability margin, and guide the supplementary control strategy. All IMs and analyses are verified by frequency scans and simulations in PSCAD.

## Nomenclature

*Abbreviations*

| | |
|---|---|
| MMC | Modular multilevel converter. |
| TL-VSC | Two-level voltage source converter. |
| BTB | Back-to-back. |
| PS/NS/ZS | Positive-/negative-/and zero-sequence. |
| DM, CM | Differential- and common-mode. |
| IM, AM | Impedance and admittance model. |
| TIM/AIM | Theoretical and analytical impedance model. |
| CCSC | circulating current suppressing control. |
| ZSCC | Zero-sequence current control. |
| SE, NE | Sending and receiving end. |
| FFT | Fast Fourier transform. |
| 1-d, 2-d | One- and two-dimensional. |
| eig, *LR* | Eigenvalues of an open-loop IM ratio. |

*Expressions*

| | |
|---|---|
| $G(x, y)$ | Element of coordinates (x, y) of $(2n+1)^{th}$ order Toeplitz matrix $G$ formed by Fourier vector $g$. |
| $(g/\Delta g)\langle i \rangle$ | $i^{th}$ harmonic component of $(2n+1)^{th}$ $(g/\Delta g)$. |
| $a_{g\langle i \rangle}+b_{g\langle i \rangle}j$ | Fourier coefficients of $g\langle i \rangle$. |
| $h(s)$ | Transfer function. |

*Parameters and Variables*

| | |
|---|---|
| $U$, $O$ | Identity and zero matrices. |
| $C$, $L$, $R$ | Capacitance, inductance, and resistance. |
| $P$, $Q$, $S$ | Active, reactive, and apparent power. |
| $Y$, $Z$ | Admittance and impedance. |
| $\theta$, $\omega$, $f$ | Angle, angular frequency, and frequency. |
| $i$, $m$, $v$ | Variable of current, insertion index, and voltage. |
| $d$, $q$ | Variable of outer loop dynamics in *dq*-frame. |
| $n$, $N$ | Number of truncated harmonic order of vectors and submodule of each arm. |
| $k$, $T$ | Unitary/proportional gain and time constant. |

*Prefix*

| | |
|---|---|
| $\Delta$ | Small-signal perturbation. |

*Subscripts*

| | |
|---|---|
| $a$, $b$, $c$ | Variable of each phase in the stationary frame. |
| $d$, $q$, $z$ | Direct-, quadrature-axis, and DC components in synchronous rotating frame. |
| b, RMS, pu | Base, root mean square, and per-unit value. |
| dc, g | DC and infinite grid terminal. |
| de, d, f, m, o | Coefficient/transfer function of decoupling, delay, feedforward, normalization, and offset. |
| v, w, a, c, ma | Transformer primary/secondary winding, turns-ratio, coupling and magnetizing coefficient. |
| p, 0, …, n | Perturbation, DC, …, n times line-frequency. |
| s, r | Sending and receiving end. |
| u, l | Upper and lower arm. |

*Superscripts*

| | |
|---|---|
| b | Base value. |
| c, d | Common- and differential-mode component. |
| p/n/z | Positive-/negative-/zero-sequence component. |
| * | Reference or rated value. |
| ′, ″ | Revised coefficient, matrix, and element. |
| 0 | Initial value for Newton–Raphson iteration. |

## I. Introduction

**M**MC-based multiterminal systems, featured by decoupled power control, good adaptability of various voltage levels, and high power quality [1], are expected to be widely applied in modern power systems. However, the recent real-world events, i.e., the 30 Hz subsynchronous oscillation [2] of the Nan-Ao wind power transmission project and the 1.2/0.7&1.8

This work was supported by the National Natural Science Foundation of China under grant U1866601. (*Corresponding author: Qirong Jiang.*)
C. Zhao and Q. Jiang are with the Department of Electrical Engineering, Tsinghua University, Beijing 100084, China (e-mail: zhaocb19@mails.tsinghua.edu.cn; qrjiang@mail.tsinghua.edu.cn).

kHz high-frequency oscillation [3], [4] of the Luxi/Yu-E grid interconnection project, attract broad concern for the system wideband small-signal stability, where the impedance-based method is welcomed for offering intuitive physical insights on the mechanism, protecting the privacy of vendors, and being able to adopt the field test.

Establishing the IM/AM that contains MMC is the basis of theoretical stability analysis, and extensive progress is reported in [2]-[20]. IM describes the ratio of the small-signal response of voltage and current at each frequency of interest. Toward a typical linear time-periodic system, IMs can be indirectly and directly derived based on the 1) harmonic state-space in the time domain [2], [4]-[10], and 2) multiharmonic linearization in the frequency domain [11]-[16]. The harmonic state-space-based method typically fails to approximate the pure time delay (cannot be omitted for MMC) above 100 $\mu$s using the rational polynomial at high frequency if starting from developing ordinary differential equations [5]-[7], [9], and even if great progress is made in [10] where the use of delay differential equations avoids the linearization of irrational transfer functions, the computation burden is high. Comparatively, the multiharmonic linearization sidesteps the flaws by eliminating intermediate variables and performing low-order matrix manipulation, and TIMs instead of AIMs that are common for TL-VSCs [17]-[19] are available (frequency responses are available for both TIM and AIM while explicit transfer functions are only available for AIM). Hence, multiharmonic linearization is more intuitive since IMs/AMs are frequency domain models and the theoretical basis of this paper.

In summary, efforts were mostly made on a single MMC in the earlier research [2]-[4], [6], [8]-[11], [16], [20], where ideal DC/AC voltage sources are deliberately assumed in general. The simplification diverges from the real world and leads to the low credibility of the obtained AC/DC IMs on stability identifying; thus urges a generalized IM framework that simultaneously considers both AC and DC terminal dynamics, which is not hitherto unified. It is also found that only the ideal transformer model is included in the current IMs and the mutual coupling is ignored, which is not considered for a long period.

In addition to the aforementioned imperfections of IMs for the single MMC, more issues are noteworthy when combining separate IMs for multiterminal stability analysis. Initially, the steady-state analytically computed in the time [21] or energy domain [22] with truncated harmonics considered, is doubtful with several simplifications, e.g., neglecting the active power loss or power flow constraints, while the measurement-based method [11] adopted in the frequency domain is invalid when systems are unstable, and the inaccuracy of steady-state impacts the IMs. Moreover, when the transformerless cases are focused in the distribution network for flexible interconnection [23], the ZS DM impedance of an MMC-BTB system is no longer infinite as same as the HVDC scenario, and the effect of virtual impedance on the stability induced by ZSCC is unclear. Finally, since the instability source can be any of the interconnections of AC and DC terminals [13], [18], [19], it is essential to analyze the stability with detailed IMs rather than simplified IMs (the complete control loops are modeled in the detailed IM while several control loops are omitted in the simplified IM) at each terminal to position the authentic unstable modes or verify some conclusions deduced from TL-VSCs, which can also help reveal the essence of some recent system-level stability analysis including MMCs [24], [25]. In other words, executing only the DC stability analysis considering AC terminal dynamics [12] or vice versa [7] is not sufficient,

After realizing the significance of facilitating the impedance-based stability analysis for complex networks with MMCs on the premise of satisfied system dynamic responses, basic works are completed by focusing on a three-phase symmetrical MMC-BTB system, which is a compromise between multiterminal and single MMC systems. This paper aims to contribute the following.

1) Since multiharmonic linearization is performed in the frequency domain, a practical frequency domain steady-state computing method is proposed based on Newton–Raphson iteration, which covers the cases of whether transformers are configured and is compared with the existing methods.

2) Based on the multiharmonic linearization, 1-d DC IMs are first obtained followed by the sequence domain 2-d AC IMs fully considering the frequency coupling, typical control loops, the pure time delay, and mutual coupling of transformers, whose accuracy is checked by frequency scans in PSCAD.

3) Aided by the derived accurate TIMs, sensitivity analyses are completed to explicate the influence on the stability of the nonideal transformer model and the virtual ZS DM impedance, and also to check the rationality of the deduced simplified IM with line-frequency NS current (suppressing) control.

4) Using the 1-d Nyquist criterion at the DC terminal and the 2-d Nyquist criterion at the AC terminal, the utility of IMs is verified by correctly positioning the instability source under multiple operating conditions. The adaptivity of a decoupled 1-d Nyquist criterion [26] and the practical criterion proposed in [27] are also examined at the AC terminal.

The remainder of this paper is organized as follows. Section II presents the elementary knowledge of the investigated system. A novel steady-state computing method is introduced in Section III followed by the derivation and validation of TIMs in Section IV. Sensitivity and stability analyses are performed in Sections V and VI, respectively. Finally, Section VII concludes the work.

## II. BASICS OF MMC-BTB SYSTEM

In this section, after introducing the configuration and control loops of the system in turn, the single-arm steady-state mathematical model is first derived in the time domain and subsequently transformed to the frequency domain, which lays the foundation of the subsequent computing method and IMs. Only the case of the ideal transformer model is focused on in this section for conciseness, and the cases of the nonideal transformer model or transformerless are comparatively discussed in Section III and Section V.

*A. Configuration*

Fig. 1 reveals the investigated MMC-BTB system based on the topology of real projects [3], [4]. Current paths and definitions of variables are briefly marked in phase *a* of SE&RE.

The adopted average model for each arm indicates that 1) the submodule capacitance-voltage that belongs to the single arm is balanced, and 2) the frequency range of concern is far below the equivalent switching frequency for stability analysis, and such assumptions are usually acceptable according to existing works. The lumped parameters are adopted for the grid side for simplicity since this paper focuses on the modeling of MMCs under various control modes.

TABLE I
KEY PARAMETERS OF THE SYSTEM

| Category | Symbol | Value (Unit) |
|---|---|---|
| Base | $S_b$, $V_{gb}$, $V_{dcb}$, $f_1/T_1$ | 10 MVA, 10 kV, 20 kV, 50 Hz/0.02 s |
| Grid | $v_{gr}$, $v_{gs}$ | 10∠81°, 11∠90° |
| | $L_{gr}$, $L_{gs}$, $R_{gr}$, $R_{gs}$ | 10 mH, 0, 0, 2 Ω |
| MMC | $L$, $R$ | 50 mH, 0.5 Ω |
| | $C'$, N | 1000 μF, 20 |
| Transformer | $k_a$, $I_{ma}$, $X_v$ | 1.25, 1%, 0.1 pu |
| Controller | $P_s^*$, $Q_s^*$, $v_{dcr}^*$, $v_{RMSr}^*$ | 5 MW, 0 MVA, 20 kV, 10 kV |
| | $h_{PLL}(s)$ | 100+1/0.05s |
| | $h_{PQ}(s)$, $h_{v_{dc}}(s)$, $h_{v_{RMS}}(s)$ | 0.1+1/0.05s, 4+1/0.05s, 2+1/0.05s |
| | $h_i(s)$, $h_{CCSC}(s)$, $h_{ZSCC}(s)$ | 0.4+1/0.01s, 0.8+1/0.01s, 50+1/0.01s |
| | $T_d$ | 300 μs |

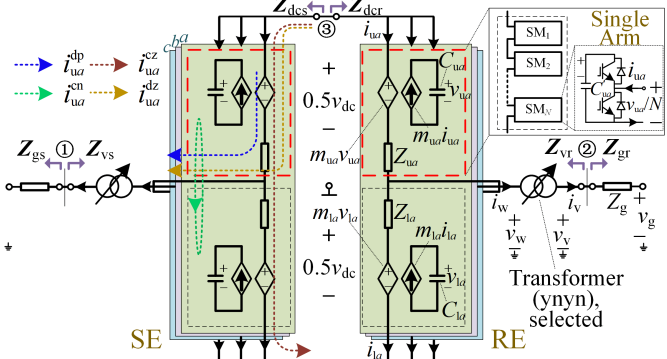

Fig. 1. Diagram of the MMC-BTB system [3], [4].

### B. Control Loops

Referring to the most common grid-following control, the loops are illustrated in Fig. 2 referring to [4], [15], where the SE regulates $P$ and $Q$ and the RE regulates $v_{dc}$ and $v_{RMS}$ for the outer loop control. NS current control and ZSCC are added in addition to the regular PS current control and CCSC for the inner loop control. $T_d$ represents the sum of control delay in the $dq$-frame while the sampling delay in the $abc$-frame is not considered, and the per-unit values are adopted for control. The key parameters are listed in Table I.

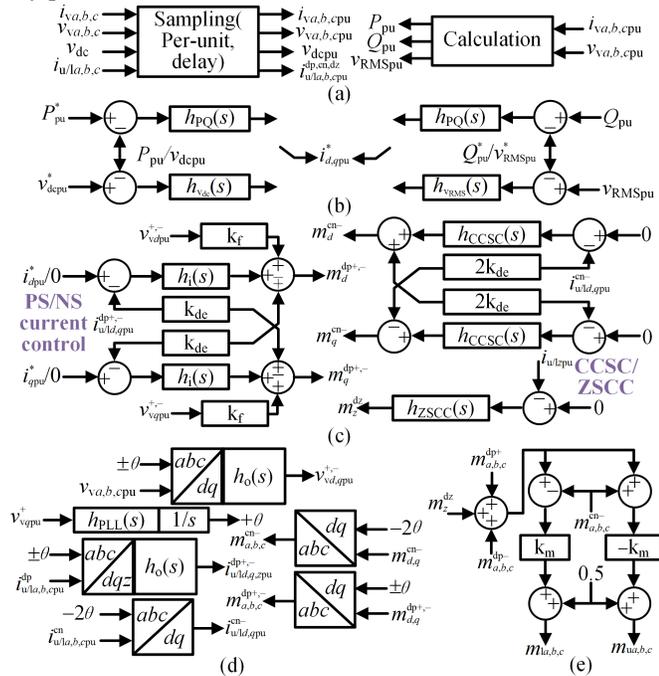

Fig. 2. Control loops. (a) Processing. (b) Outer loop control. (c) Inner loop control. (d) (Inverse) Park transformation. (e) Insertion index generation.

### C. Single-arm Time Domain Model

For the IM of a single MMC, by grasping the symmetric and periodic distribution of the internal multiharmonic variables, [11] introduces a focus on the single arm instead of the single phase [2], [6], [9]. Thus, this idea is extended to the BTB system. The dynamics of the upper arm, i.e., phase $a$ of RE, can be reflected in the time domain:

$$\begin{cases} L(di/dt) + Ri = 0.5v_{dc} - v_w - mv - v^{dz} \\ C(dv/dt) = mi, \; C = C'/N \end{cases} \quad (1)$$

where the subscript "$ua$" is neglected for simplification unless specified. $v^{dz}$ is the voltage between the grounding and neutral point and acts as the source of ZS DM circuit. An extra equation of the ideal transformer model is:

$$v_w = [v_v + L_v(d(2i_{ua}^{dp}/k_a)/dt)]/k_a \quad (2)$$

Substituting (2) to (1) eliminates $v_w$ but retains $v_v$ and $i$, which matches the demands of the voltage sampling in Fig. 2 and the arm current decomposition of deriving IMs.

### D. Frequency Domain Interpretation

By replacing the time domain variable multiplication with the vector convolution and further matrix-vector multiplication both in the frequency domain [11], (1) and (2) are converted to:

$$\begin{cases} \boldsymbol{i} = \boldsymbol{Y}_{LR}(0.5\boldsymbol{v}_{dc} - \boldsymbol{v}_v/k_a - \boldsymbol{M}\boldsymbol{v} - \boldsymbol{v}^{dz}) \\ \boldsymbol{Y}_C \boldsymbol{v} = \boldsymbol{M}\boldsymbol{i} \end{cases} \quad (3)$$

$\boldsymbol{Y}_{LR}$ and $\boldsymbol{Y}_C$ are diagonal transfer function matrices:

$$\begin{aligned} \boldsymbol{Y}_{LR} &= \text{diag}\{[s+j(x-1-n)\omega_1]L+R\}^{-1} \\ &+ \text{diag}\{h \cdot 2/k_a^2 \cdot 0.5(1+g)\{[s+j(x-1-n)\omega_1]L_v\}\}^{-1}, \\ \boldsymbol{Y}_C &= \text{diag}\{[s+j(x-1-n)\omega_1]C\}, \; 1 \le x \le 2n+1, \\ h &= \text{sign}\{\text{mod}[(x+2-n),6]\}, \; g = 2\text{mod}[(x-1-n),2]-1. \end{aligned} \quad (4)$$

Hence, $\boldsymbol{v}_v$, $\boldsymbol{m}$, $\boldsymbol{i}$, and $\boldsymbol{v}$ are the steady-state vectors to be solved and form the corresponding Toeplitz matrices in (3). $\boldsymbol{v}^{dz}$ induces the coupling of both ends, which differs from the single MMC:

$$\boldsymbol{v}^{dz} = -\sum_{t=s,r} \frac{[\boldsymbol{Y}_{LRt} + 2/k_a^2 \cdot \boldsymbol{Y}_{gt}](x,x) \cdot (\boldsymbol{Mv})_t \langle \pm 3 \rangle}{\sum_{t=s,r}[\boldsymbol{Y}_{LRt} + 2/k_a^2 \cdot \boldsymbol{Y}_{gt}](x,x)}, \; x = n+1 \pm 3. \quad (5)$$

Due to the transformer, h in (4) guarantees the admittance of ZS DM circuit to be zero, and $\boldsymbol{v}^{dz}$ can be removed from (3), so ZS decouples from PS or NS in the DM circuit. Such coincidence should be assessed in the transformerless case.

## III. FREQUENCY DOMAIN STEADY-STATE COMPUTING METHOD OF MMC-BTB SYSTEM

The steady-state of TL-VSC can be analytically obtained based on the intuitive physical meaning but is unrealistic for the

MMC considering its multiharmonic feature. A numerical method (b) is proposed and compared with the measurement-based method (a) and simulation (c, as the standard).

*A. Idea and Hypotheses*

Solving the phase and amplitude of each order of harmonics for a time domain variable can be considered as solving $a_{g\langle i\rangle}$ and $b_{g\langle i\rangle}$ of frequency domain vectors with $s=j\omega_1(2\pi f_1), j\omega_2, \ldots, j\omega_n$. Several hypotheses are formulated as follows:

1) Supposing that the steady errors are neglected, the control objectives of both inner and outer loops are equal to their references, e.g., $v_{dc}=v_{dc}^*=v_{dcb}$, thus $v_{dc}$ becomes a known quantity.

2) Due to the conjugate relation between the $(n+1\pm i)^{th}$ elements of each vector, $\boldsymbol{v}$ introduces $(2n+1)$ unknowns; the use of CCSC and ZSCC/transformer lead $\boldsymbol{i}\langle 2\rangle$ and $\boldsymbol{i}\langle 3\rangle$ to be 0, hence $\boldsymbol{i}$ introduces $(2n-3)$ unknowns; based on Fig.2, $\boldsymbol{m}\langle 0\rangle=0.5$ while $\boldsymbol{m}\langle 1\rangle$ (led by PS/NS current control), $\boldsymbol{m}\langle 2\rangle$ (led by CCSC), and $\boldsymbol{m}\langle 3\rangle$ (led by ZSCC) can be nonzero; $\boldsymbol{v}_v\langle 1\rangle$ is related with the AC power flow and introduces two more unknowns. Hence, at most $(4n+6)$ equations are required for each end.

3) The iteration between $\boldsymbol{v}$ and $\boldsymbol{i}$ in (3) requires the truncation of each vector. Considering $\boldsymbol{m}\langle 3\rangle$ is specially focused, n is selected as 4 in this paper.

*B. Process and Annotations*

A flow chart is offered to explain the process by first computing coefficients of SE and subsequently RE, as shown in Fig. 3, and annotations are added as follows.

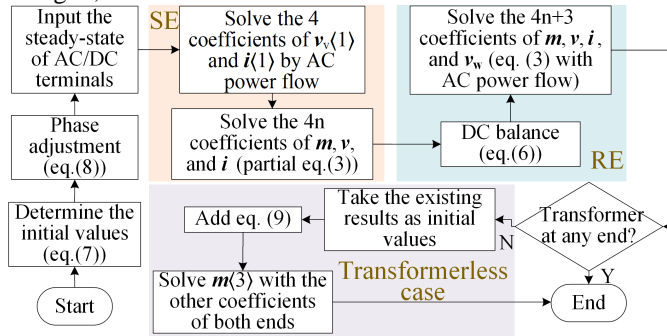

Fig. 3. Flow chart of the frequency domain steady-state computing method.

1) With the given $P^*$, $Q^*$, $Z_g$, and $v_g$, 4 coefficients contributed by $\boldsymbol{v}_v\langle 1\rangle$ and $\boldsymbol{i}\langle 1\rangle$ can be independently solved based on the AC power flow for SE. Suppose that the transformer is first configured and $\boldsymbol{m}\langle 3\rangle=0$; then, the remaining $4n$ coefficients of SE contributed by $\boldsymbol{m}$, $\boldsymbol{v}$, and $\boldsymbol{i}$ can be independently solved with the same number of real equations featured by the balance of $a_{g\langle i\rangle}$ and $b_{g\langle i\rangle}$ extracted from the matrix form of (3).

2) The DC balance holds for SE and RE:

$$v_{dcr}\langle 0\rangle = v_{dcs}\langle 0\rangle, \ \boldsymbol{i}_r\langle 0\rangle = -\boldsymbol{i}_s\langle 0\rangle. \quad (6)$$

Eq. (6) satisfies because of the negligible electrical distance between the DC terminals in the grid interconnection scenario and can be adjusted if the voltage drop of DC line impedance must be considered in other scenarios. It is also confirmed for RE that the remaining $4n+3$ coefficients should be simultaneously solved using the $4n$ equations from (3), two equations to represent the voltage drop of $Z_{gr}$, and an extra equation that represents the RMS voltage control.

3) The aforementioned nonlinear algebraic equations can be numerally solved by Newton–Raphson iteration, and part of the initial values are given as (the rest are default to 0):

$$a^0_{\boldsymbol{v}_v\langle 1\rangle}=\sqrt{2}V_b/2\sqrt{3},\ a^0_{\boldsymbol{m}\langle 1\rangle}=-2a^0_{\boldsymbol{v}_1\langle 1\rangle}/V_{dcb},\ a^0_{\boldsymbol{v}\langle 1\rangle}=V_{dcb}. \quad (7)$$

4) When handling each coefficient for the calculation, the phase adjustment relative to $\boldsymbol{v}_v\langle 1\rangle$ is necessary:

$$a'_{\boldsymbol{g}\langle i\rangle}=\text{abs}(\boldsymbol{g}\langle i\rangle)\cos[\text{ang}(\boldsymbol{g}\langle i\rangle)-\text{ang}(\boldsymbol{v}_v\langle 1\rangle)],$$
$$b'_{\boldsymbol{g}\langle i\rangle}=\text{abs}(\boldsymbol{g}\langle i\rangle)\sin[\text{ang}(\boldsymbol{g}\langle i\rangle)-\text{ang}(\boldsymbol{v}_v\langle 1\rangle)], \quad (8)$$
$$\text{abs}(\cdot)=\sqrt{a_{(\cdot)}^2+b_{(\cdot)}^2},\ \text{ang}(\cdot)=\arctan(b_{(\cdot)}/a_{(\cdot)}).$$

5) If the transformerless case is focused on, these results can serve as the initials added with $a^0_{\boldsymbol{m}\langle 3\rangle}=0$, $b^0_{\boldsymbol{m}\langle 3\rangle}=0$. With identical ZS circuit sampled and ZSCC for both ends, the follow-up (9) should be combined with the aforementioned $8n+3$ equations extracted from (3) for a simultaneous solution:

$$(\boldsymbol{Mv})_r\langle 3\rangle-(\boldsymbol{Mv})_s\langle 3\rangle=0. \quad (9)$$

*C. Validation*

The steady-state of the transformerless case derived by three methods towards the operating condition in Table I is listed and compared in Table II and Fig. 4, respectively, as an example. The real and imaginary parts of each coefficient are listed in the upper and lower roles of each space in Table II.

TABLE II
COMPARISON BETWEEN CALCULATING METHODS OF THE SYSTEM STEADY-STATE (UPPER ARM, PHASE A, TRANSFORMERLESS CASE)

| Variable | RE method a | RE method b | RE method c | SE method a | SE method b | SE method c |
|---|---|---|---|---|---|---|
| $v_v\langle 1\rangle$ | 10.00 0.00j | 10.00 0.00j | 10.00 0.00j | 10.00 0.00j | 10.00 0.00j | 10.00 0.00j |
| $m\langle 1\rangle$ | –0.21 –0.050j | –0.21 –0.050j | –0.21 –0.050j | –0.20 0.052j | –0.20 0.052j | –0.20 0.052j |
| $m\langle 2\rangle$ | –0.0064 0.037j | –0.0072 0.037j | –0.0064 0.037j | –0.0082 –0.037j | –0.0079 –0.037j | –0.0082 –0.037j |
| $m\langle 3\rangle$ | 0(e$^{-4}$) –0.026j | 0(e$^{-4}$) –0.025j | 0(e$^{-4}$) –0.026j | 0(e$^{-4}$) –0.026j | 0(e$^{-4}$) –0.025j | 0(e$^{-4}$) –0.026j |
| $i\langle 0\rangle$ | 0.099 | 0.082 | 0.082 | –0.069 | –0.082 | –0.082 |
| $i\langle 1\rangle$ | 0.12 0.012j | 0.099 0j | 0.099 0j(e$^{-4}$) | –0.086 –0.016j | –0.10 0j | –0.10 0j(e$^{-4}$) |
| $i\langle 2\rangle$ | **0.021 0.016j** | 0 0j | 0(e$^{-6}$) 0j(e$^{-6}$) | **0.022 –0.017j** | 0 0j | 0(e$^{-6}$) 0j(e$^{-6}$) |
| $i\langle 3\rangle$ | **0(e$^{-4}$) –0.0064j** | 0 0j | 0(e$^{-5}$) 0j(e$^{-5}$) | **0(e$^{-4}$) 0.0064j** | 0 0j | 0(e$^{-5}$) 0j(e$^{-5}$) |
| $i\langle 4\rangle$ | –0.044 –0.022j | –0.0061 0.0010j | –0.0057 0j(e$^{-4}$) | –0.031 –0.019j | –0.0045 –0.0031j | –0.0048 –0.0033j |
| $v\langle 0\rangle$ | 20.55 | 20.44 | 20.44 | 20.46 | 20.66 | 20.66 |
| $v\langle 1\rangle$ | 0.17 –2.17j | 0.025 –2.02j | 0.031 –2.02j | –0.20 2.06j | 0.029 2.15j | 0.029 2.15j |
| $v\langle 2\rangle$ | 0.25 0.46j | 0.075 0.67j | 0.068 0.67j | –0.090 –0.88j | 0.088 –0.67j | 0.10 –0.67j |
| $v\langle 3\rangle$ | –0.022 –0.050j | –0.057 –0.0050j | –0.057 –0.0053j | 0.081 –0.068j | –0.047 –0.063j | –0.047 –0.061j |
| $v\langle 4\rangle$ | –0.22 0.45j | 0(e$^{-4}$) 0.047j | 0(e$^{-4}$) 0.047j | –0.19 0.30j | 0.031 –0.025j | 0.032 –0.025j |

In [11], $\boldsymbol{m}$ is first obtained by simulation and then brought into (3) to calculate $\boldsymbol{i}$ and $\boldsymbol{v}$ for method a, but obvious errors are observed especially $\boldsymbol{i}\langle 2\rangle$ (should be 0 theoretically due to CCSC) and $\boldsymbol{i}\langle 0\rangle$ (as (6) indicates). Particularly, since a steady $\boldsymbol{m}$ cannot be measured when the system is unstable, neither method a nor method c is rigorous, while the proposed method b has high precision and low computational overhead (normally, 4-6 iterations are adequate with proper initial values as given (7)), and avoids the complicated but unnecessary transformation from the time or energy domain to the frequency domain [21], [22]. The magnitude of $m\langle 3\rangle$ is nearly 70% of that of $m\langle 2\rangle$ due to the nonexcessive $L$ and $C$ in the transformerless case that is possible for distribution networks. Considering the critical impact of CCSC on the stability [11], [12], it can be referred ZSCC cannot be neglected in the IM if it works and will be elaborated in Section V.

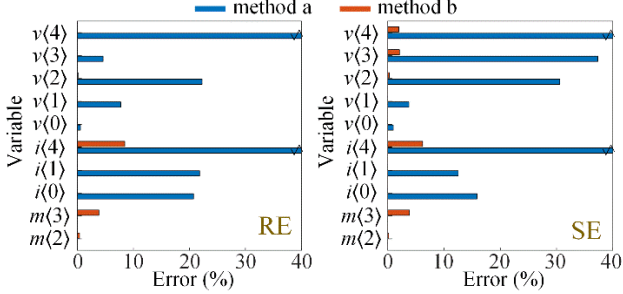

Fig. 4. Amplitude error of methods a and b compared with method c.

## IV. DERIVATION OF AC/DC IMs OF THE MMC-BTB SYSTEM

The core work of developing IMs is introduced in this section. The modeling idea is clarified using the power stage and efforts are added to the control stage, whose transfer function matrices are given in the Appendix for readability. DC and AC TIMs are then obtained and verified.

### A. Basic Modeling Idea

There is a duality between AC and DC IMs but was not explicitly announced. Multiharmonic linearize on (3) gives:

$$\begin{cases} \Delta i = Y_{LR}(0.5\Delta v_{dc} - \Delta v_v / k_a - M\Delta v - V\Delta m) \\ Y_C \Delta v = M\Delta i + I\Delta m \end{cases} \quad (10)$$

$\Delta m$ is closely related to the sampling and control as illustrated in Fig. 2:

$$\Delta m = B_i \Delta i + B_{v_v} \Delta v_v + B_{v_{dc}} \Delta v_{dc} \quad (11)$$

It is worth noting that $\Delta v$ does not contribute to $\Delta m$ as Fig. 2 reveals, but if energy-based control generates the reference of CCSC, an extra $B_v$ should be developed in (11). Since IM describes the relation between selected elements of $\Delta i$ with $\Delta v_v$ or $\Delta v_{dc}$, $\Delta m$ and $\Delta v$ are the intermediate variables and can be eliminated by combining (10) and (11). In addition, an extra equation to eliminate $\Delta v_v$ or $\Delta v_{dc}$ for DC or AC IM is available since the dynamics can be described from the two sides of each partition point in Fig. 1 based on Kirchhoff's voltage law. Due to the simpler realization of DC IMs, they are first derived and subsequently used for deriving AC IMs.

Furthermore, two points are claimed: 1) AMs should be first developed followed by the inversion to IMs and 2) since $s=j\omega_p$ of the IM corresponds to $s=j\omega_1$ of the steady-state, there is a position offset between two elements in $(\Delta g)\langle i\rangle$ and $(g)\langle i\rangle$ that hold the same CM-DM/sequence distribution law [11].

### B. Control Stage

Eq. (11) is transformed to (12) for the modeling modularity:

$$-\Delta m = J\Delta i + K\Delta v_v + E\Delta d + F\Delta q \quad (12)$$

where $J$ and $K$ describe the influence of the inner loop control on $\Delta m$ from $\Delta i$ and $\Delta v_v$ with the fixed reference values; $\Delta d$ and $\Delta q$ cooperate with $E$ and $F$ to reflect the influence of the outer loop control on $\Delta m$ without the inner loop control.

Since $\Delta v_v=0$ ensures $\Delta \theta=0$, a process of current sampling, (inverse) Park transformation, and linear control ensures $Q$ a diagonal matrix. $P$ is related to the feedforward and phase-locked loop (PLL), where the former is similar to $Q$ and the latter introduces off-diagonal terms. The closed-loop PLL transfer function considering multiple delays reflected by $h_o$ ($T_1/4$ phase-shift offset) and $h_d$ (aggregated control delay) is:

$$h_{PLL}(s) = \frac{\Delta\theta}{\Delta v_v \langle \pm 1\rangle} = e^{-\text{ang}(v_v\langle 1\rangle)}[h_{PLL}'(s) + 2\text{abs}(v_v\langle 1\rangle)]$$

$$h_{PLL}'(s+j\omega_1) = [k_v \cdot [h_o \cdot h_d](s) \cdot h_{PLL}(s) \cdot (1/s)]^{-1} \quad (13)$$

$\Delta d$ and $\Delta q$ depend on the control modes, whose relations with the sampling are described by matrices $U$, $R$, $S$, $T$, $W$, and $V$. The cascade controllers are contained in $E$ and $F$. Finally, $B_i$, $B_{v_{dc}}$, and $B_{v_v}$ can be expressed with matrices in this subsection.

### C. Derivation and Validation of DC IM

Since $i_{dc}$ is 3 times $i_{ua}^{cz}$ while $i_w$ is 2 times $i_{ua}^{dp}$, the extra equation to eliminate $\Delta v_v$ for the 1-d DC AM/IM $Y_{dc}/Z_{dc}$ is:

$$\Delta v_v = 2/k_a \cdot \underbrace{0.5(1+g)\text{diag}[(s+j(x-1-n)\omega_1)L_g]}_{Z_g(s)(x,x)}\Delta i, \; x = n\pm 1 \quad (14)$$

Bringing (11) and (14) into (10) gives:

$$Y_{dc} = (3A_{dc}^{-1}B_{dc})_{[(n,n)]}, \; Z_{dc} = Y_{dc}^{-1}, \; B_{dc} = 0.5Y_C Y_{LR} - BB_{v_{dc}},$$

$$A_{dc} = A + BB_i + 2/k_a \cdot Z_g (BB_{v_v} + Y_C Y_{LR}), \quad (15)$$

$$A = Y_C + Y_{LR}M^2, \; B = Y_{LR}(MI + Y_C V).$$

The accuracy of DC IMs of both ends is validated as shown in Fig. 5. The solid lines represent the theoretical IM by taking frequency of interest into (15) while the marks represent the frequency scans. It manifests the feature of introducing passivity only for RE at the high-frequency as the zoom-in shows due to the sampling delay of $v_{dc}$.

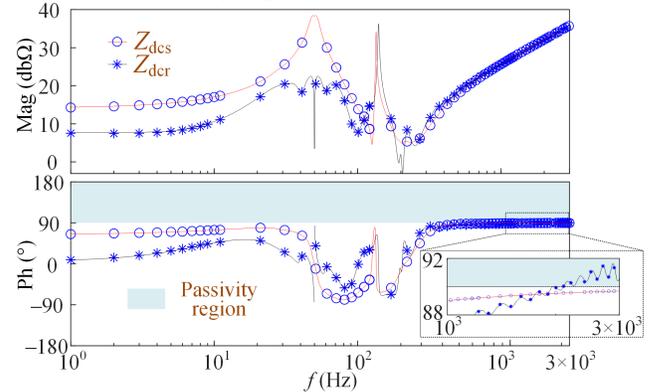

Fig. 5. Validation of DC IMs.

### D. Derivation and Validation of AC IM

To derive the sequence domain 2-d AC AM/IM $Y_v/Z_v$, the following dual expression of (14) to eliminate $\Delta v_{dc}$ is:

$$\Delta v_{dc} = -3\hat{Z}_{dc}\Delta i \quad (16)$$

where "^" indicates that only the subscript s or r of $Z_{dc}$ is inconsistent with that of the other matrices in (17):

$$Y_v = (2/k_a \cdot A_v^{-1} B_v)\begin{bmatrix}(n+1,n+1) & (n+1,n-1) \\ (n-1,n+1) & (n-1,n-1)\end{bmatrix}, Z_v = Y_v^{-1},$$

$$B_v = (Y_C Y_{LR})/k_a + BB_{v_v}, \quad (17)$$

$$A_v = A + BB_i + (0.5 Y_C Y_{LR} - BB_{v_{dc}})\hat{Z}_{dc}/3.$$

With the conjugate symmetry of 2-d $Z_{ac}$ [5], only the two elements that belong to the same row are plotted for each end in Fig. 6, and the modeling accuracy is confirmed similar to Fig. 5. Since matrices $A$, $B$, $B_i$, $B_{v_{dc}}$, and $B_{v_v}$ are shared in (15) and (17), there must be a generalized modeling framework for AC and DC IMs.

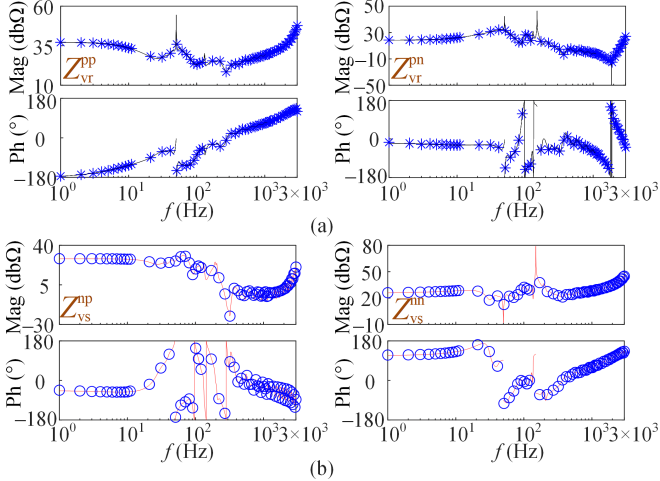

Fig. 6. Validation of the AC IMs. (a) RE. (b) SE.

## V. SENSITIVITY ANALYSIS

Since the derived IMs are based on (3) of the ideal transformer model in Section IV, the influence on IM of the nonideal transformer model and the case without transformers remain unclear. Detailed IMs can also serve as standards to check the errors of simplified IMs at high frequencies. Due to the pure time delay and the inverse of $A_v$ or $A_{dc}$ in AMs, AIMs are impractical, and the sensitivity analyses are completed by adjusting key parameters and replotting TIMs point-by-point.

### A. Issues of Nonideal Transformer Model

The classical modeling approach of the nonideal transformer is based on the following frequency domain formula in PSCAD:

$$\begin{bmatrix} v_v\langle\pm 1\rangle \\ v_w\langle\pm 1\rangle \end{bmatrix} = s\begin{bmatrix} L_{vv} & L_{vw} \\ L_{vw} & L_{ww} \end{bmatrix}\begin{bmatrix} -i_v\langle\pm 1\rangle \\ i_w\langle\pm 1\rangle \end{bmatrix} \quad (18)$$

where $L_{vv}$, $L_{ww}$, and $L_{vw}$ are the self-inductances of winding v/w and the mutual inductance. Thus, (2) of the ideal transformer model turns to (19) of the nonideal transformer model:

$$v_v\langle\pm 1\rangle = (L_{vw}/L_{ww})\cdot v_w\langle\pm 1\rangle - sL_{vv}(1-k_c^2)\cdot 2i\langle\pm 1\rangle/k_a,$$

$$2i\langle\pm 1\rangle/k_a = (L_{vv}/L_{vw})\cdot i_v\langle\pm 1\rangle - v_v\langle\pm 1\rangle/(sL_{vw}), \quad (19)$$

$$k_c = L_{vw}/\sqrt{L_{vv}L_{ww}}.$$

By calculating with $k_a$, $I_{ma}$, $X_v$, and $S_b$ in Table I, $k_c$ is 0.9995 while both $(L_{vv}/L_{vw})$ and $(L_{vw}/L_{ww})$ are close to $k_a$. Generally, the coupling between $i_v$ and $v_v$ in the second equation of (19) induces an extra negative impedance $Z_{vv}$ shunt connected with the deduced $Z_{ac}$ in (17), which mainly affects the low-frequency impedance characteristics of the AC IM and doubts the simple handling in [12]. However, DC IM is almost unaffected since $Z_{vv}$ is far greater than $Z_g$. Such inferences are supported as shown in Fig. 7, where the steady-state is recalculated considering (19), and the appropriate adjustments are given on control stages, e.g., replacing $k_a$ to $(L_{vv}/L_{vw})$ and updating $B_{v_v}$ for both AC and DC IMs while further revising (17) for DC IMs.

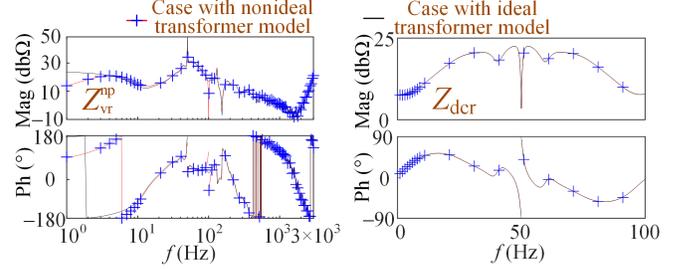

Fig. 7. Influence on AC/DC IM of the nonideal transformer.

### B. Issues of Transformerless Case

Here the doubt of the transformerless case on decoupling ZS with PS/NS DM stability analysis is removed, where ZSCC is activated and an extra submatrix is introduced for $J$:

$$J_3^z(s)(x,x) = -0.5(1+g)\cdot h_d(s+j\omega_1)$$
$$\cdot [h_{ZSCC}(s+j(x-3g)\omega_1)]\cdot 2k_m k_i, \quad x = n\pm 3. \quad (20)$$

With the computation of $m\langle 3\rangle$ in (9), the principle of $i_{ua}^{dz}=0$ and $v^{dz}$ elimination is clarified in Fig. 8 (a). The sufficient virtual impedance balance $Mv\langle 3\rangle$ with its small-signal $(M\Delta v+V\Delta m)\langle 3\rangle$ of two ends over the wideband by $h_{ZSCC}(s)$ and the idea of h in (4) can be moved from $Y_{LR}$ to its product terms in $A$ and $B$. Hence, two more effects on IMs induced by ZSCC are: 1) the change of $M$ as Table II shows and 2) the release of $Y_C(x,x)$, $x=n+1\pm 3$ in $A_v$, $B_v$, $A_{dc}$, and $B_{dc}$.

A 1-d positive sequence $Z_v^p$ is transformed from 2-d $Z_v$ and $Z_g$ for a more intuitive analysis [26]:

$$Z_v^p = Z_v^{pp} - Z_v^{np}Z_v^{pn}/(Z_g^{nn}+Z_v^{nn}) \quad (21)$$

By comparing the case transformer configured ($k_a=1$, $X_v=0$) for isolation and the transformerless case in Fig. 8 (b), a larger $k_{pZSCC}$ damps the resonant peaks over $3f_1$ to $5f_1$ and has a cumulative effect at low-frequency according to (3). In addition, the influence on the stability of 3rd harmonic injection to suppress the fluctuation of $v$ [28] remains an open issue based on the conclusion of this subsection and is worth investigating.

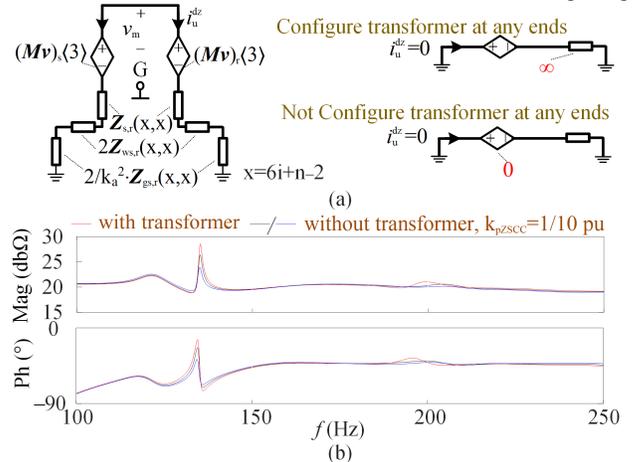

Fig. 8. Influence on AC/DC IM of whether any transformer configured. (a) Diagrams of the principle of $i_{ua}^{dz}=0$ for various cases. (b) 1-d PS Bode plots.

## C. Issues of High-frequency Simplified IMs

PS AIMs are welcomed to design notch-filter-based or self-adaptive high-frequency oscillation suppressing control [4], [16], [20]. The proposed simplified AIM is given in (22) at the bottom of this page with the following annotations:

1) With the limited bandwidth of outer-loop control, only the inner-loop current control and voltage feedforward are included, which is a cognition for high-frequency simplified IM. The coefficients k should be elaborately managed based on the control loops in Fig. 2.

2) Distinguishing from the PS control that an $f_p/f_n$ ($f_n=f_p–2f_1$) perturbation in the sequence domain results in an $f_p–f_1/f_n+f_1$ perturbation in the $dq$ domain through park transformation and then yielding an $f_p/f_n$ response through inverse park transformation, the effect of NS control is that an $f_p/f_n$ perturbation in the sequence domain results in an $f_p+f_1/f_n–f_1$ perturbation in the $dq$ domain and finally an $f_p/f_n$ response in the sequence domain, so the NS control should be fully modeled in (22) but is not considered for real project analysis in [4] or the related works in [15]-[20].

3) Another critical observation is that $v\langle 0\rangle$ replaces $v_{dc}\langle 0\rangle$ since MMC uses $v$ rather than $v_{dc}$ for modulation, and it proves the necessity of steady-state calculation again in Section III.

Rationality check of the simplified AC IMs over the high-frequency is reflected in Fig. 9. Whether NS current control is modeled or not in (22), nearly 20-Hz passivity region deviation or unacceptable errors due to the $T_1/4$ phase offset are observed, as the blue or red solid lines show. It is also confirmed that the 4th-order Padé approximation on a 300 $\mu$s pure time delay at multiple sampling paths is most effective below 200 Hz, which casts doubts on the rational approximation when establishing TIMs considering time delay. Since the frequency responses are of practical requirements for stability analysis, keeping the preliminary irrational from $e^{–sT}$ is the key point of correctly treating the time delay, as shown in this paper and [10].

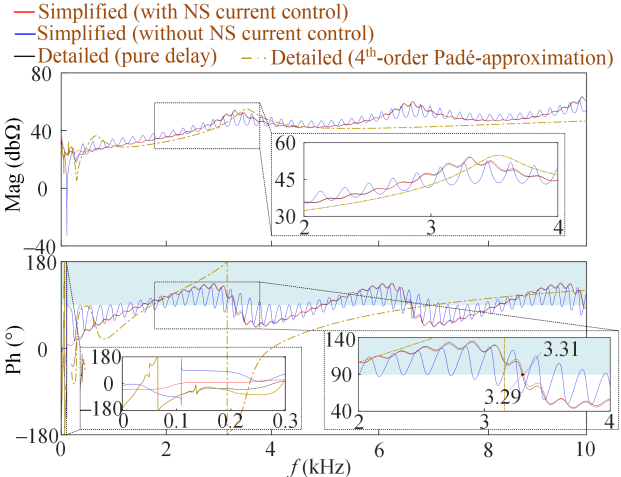

Fig. 9. Comparisons of AC IMs over the high-frequency.

## VI. CASE STUDY

Three cases are studied in this section: the change in short-circuit ratio (a), the parameter tuning of $h_{v_{dc}}(s)$ (b), and the step of $P$ (c), are discussed. Since some cases may not be close to those of the real world and the overhead lines are replaced by a lumped $Z_g$ in particular, the studies mostly verify the necessity and wideband accuracy of the TIMs. The simulation parameters are based on Table I without specification.

An alternative method to determine whether there is a right-half plane pole in the transfer function matrix ***LR*** is to examine whether the subsystems in Fig. 10 can stably operate through simulation. The ***LR*** of each terminal in Fig. 1 is defined as:

$$\begin{aligned}
&① – \text{SE, AC(2 - d)}: \boldsymbol{LR}_{vs} = \boldsymbol{Z}_{gs}\boldsymbol{Y}_{vs}, \\
&② – \text{RE, AC(2 - d)}: \boldsymbol{LR}_{vr} = \boldsymbol{Z}_{gr}\boldsymbol{Y}_{vr}, \quad (23)\\
&③ – \text{DC(1 - d)}: \boldsymbol{LR}_{dc} = \boldsymbol{Z}_{dcr}\boldsymbol{Y}_{dcs}.
\end{aligned}$$

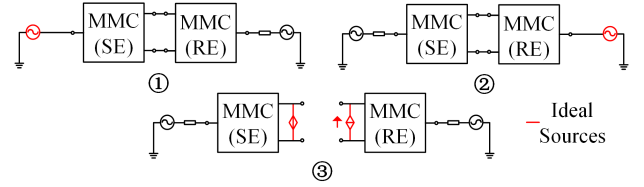

Fig. 10. Definition of subsystems for stability analyses at various terminals.

This method is acceptable in engineering since TIMs cannot directly offer system zeros and poles of MMC, which is emphasized to correctly apply Nyquist criteria with frequency responses. Considering the conjugate symmetry of $\text{eig}_p$ and $\text{eig}_n$ for AC subsystems, and the self-symmetry of $\text{eig}_{dc}$ for DC subsystem, the symmetries are both about $f_1$ for the sequence domain model, –2900 Hz to 3000 Hz of $\text{eig}_p$ and 50 Hz to 3000 Hz of $\text{eig}_{dc}$ with a step of 0.1 Hz are selected for the eigenvalue calculations.

### A. Case A: Change of Short-Circuit Ratio

If the nonideal instead of the ideal transformer model is adopted and $L_{gr}$ is increased from 22 mH to 25 mH (short-circuit ratio from 1.45 to 1.27), a diverged oscillation is first observed at $t=4$ s and subsequently becomes a sustained oscillation for all of $i_{vra}$, $i_{vsa}$ (the harmonic amplitude is very small), and $i_{dc}$ in Fig. 11. The FFT shows that $|f_p–f_1|=46.5$ Hz. Only $\text{eig}_{pr}$ of the nonideal transformer model judges an instability mode as shown in Fig. 12, whose model frequency (the initial oscillating frequency, a deviation with the sustained oscillation frequency is reasonable) and PM are approximately 98.2 Hz ($|f_p–f_1|=48.2$ Hz) and –0.3°, respectively, aided by the equivalent 1-d PS Bode plot, as Fig. 13 shows.

The above theoretical analyses are consistent with the simulations with two additional remarks on case a: 1) using the ideal transformer model to deduce IM may omit the sub/super synchronous instability mode; 2) since the instability mode is only correctly identified at RE AC instead of SE AC or the DC terminal, the supplementary control should be added in the control loops of RE related to its pure AC dynamics for priority such as PLL or inner-loop current control instead of the DC voltage control, even if DC voltage control also influence the AC sub-super synchronous dynamic responses [7].

$$Z_v^{p\prime} = \frac{k_a\{\underbrace{\boldsymbol{Z}_{LR}(n+2,n+2)}_{\text{Physical impedance}} + 2v\langle 0\rangle\cdot k_m k_i\cdot\underbrace{\{[h_i(s) - jk_{de}]\cdot[h_o\cdot h_d](s)\}}_{\text{PS current control}} + \underbrace{[h_i(s+j2\omega_1) + jk_{de}]\cdot[h_o\cdot h_d](s+j2\omega_1)\}}_{\text{NS current control}}\}}{2\{1/k_a - v\langle 0\rangle\cdot k_m k_f k_v\cdot\underbrace{\{[h_o\cdot h_d](s) + [h_o\cdot h_d](s+j2\omega_1)\}}_{\text{PS/NS voltage feedforward}}\}} \quad (22)$$

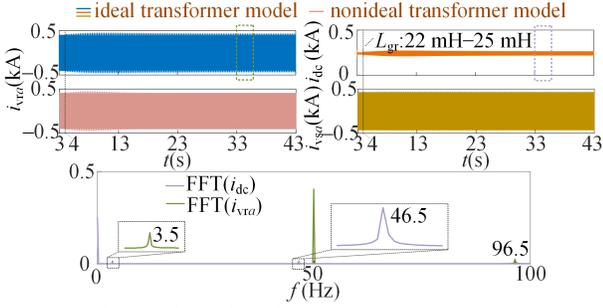

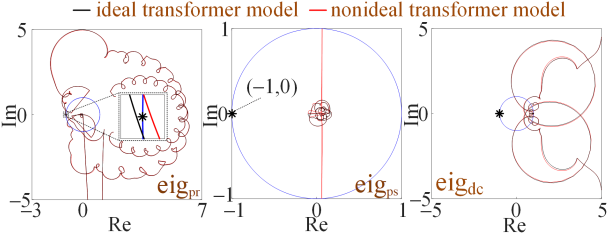

Fig. 11. Time domain simulation with the FFTs (case a).

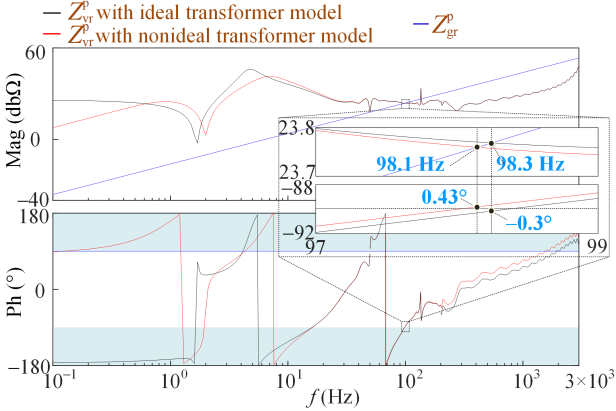

Fig. 12. 2-d Nyquist plots (case a).

Fig. 13. 1-d PS Bode plots (case a).

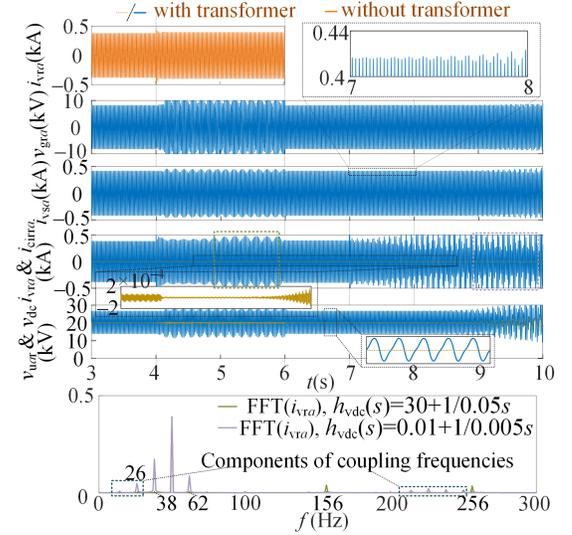

Fig. 14. Time domain simulation with the FFTs (case b).

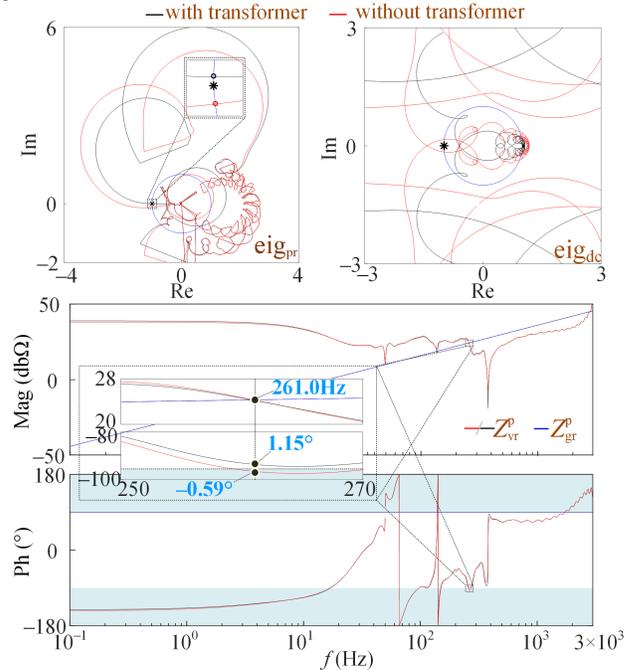

Fig. 15. Nyquist and Bode Plots ($h_{vdc}(s)=30+1/0.05s$).

## B. Case B: Tuning of $H_{vdc}(s)$

To ensure the universality of conclusions in subsection A, case b is studied, where the leakage inductance of transformer is deliberately omitted to examine the necessity of Section V, subsection B. A group of heuristic simulations is given for each subsystem, which has the time logic as follows:

1) At $t=3$ s, $L_{gr}$ is set to 0 and $k_{pvdc}$ is set to 30; 2) At $t=4$ s, $L_{gr}$ becomes its rated value $L_{gr}^*$ (10 mH); 3) At $t=6$ s, $k_{pvdc}$ becomes 4; 4) At $t=7$ s, $k_{pvdc}$ becomes 0.01, $T_{ivdc}$ becomes 0.005, and $L_{gr}$ becomes 0 again; 5) At $t=8$ s, $L_{gr}$ becomes $L_g^*$. AC/DC terminal and arm capacitor voltages as well as the phase/circulating currents are shown in Fig. 14, and further simulations prove that subsystem ② can stably operate whether the transformers are configured when $h_{vdc}(s)=30+1/0.05s$ instead of $h_{vdc}(s)=0.01+1/0.005s$, while subsystem ① cannot stably operate with any controllers and subsystem ③ can stably operate with both controllers (the simulation is neglected).

Furthermore, two instability cases are identified by $eig_{dc}$ when $L_{gr}^*$ is accessed to RE with transformers configured in Fig 15 and 16, whose corresponding Bode plots match the FFTs in Fig. 14 ($|f_p–f_1|$=206 Hz & 12 Hz). $eig_{dc}$ also confirms the stable operation of the transformerless case with a PM of 1.15° when $h_{vdc}(s)=30+1/0.05s$. However, AC IMs cannot determine the instabilities that are induced by the improper $h_{vdc}$ except $eig_{pr}$ in Fig. 15 (two $eig_p$s are omitted and they cannot determine the instabilities).

The remarks on case b are: 1) the instabilities induced by the too large or too small $k_{pvdc}$ regard to different system modes, and detailed IMs should be used to correctly identify them; 2) the discussion in Section V, subsection B is necessary, and the effect of ZSCC on the stability over $3f_1$ to $5f_1$ should be faithfully modeled; 3) in subsection A, the instability mode is only identified at the RE AC terminal, whereas part of cases of $k_{pvdc}$ tuning can be analyzed at both AC and DC terminals of the same end (RE) but is no longer feasible for the AC SE terminal; thus, the independent black-box analyses are invalid for the (inter)harmonics of $i_{vsa}$, which should be explained as a phenomenon of harmonic balance, thus urging an "impedance interaction" between multiterminal for stability modeling and instability positioning in future privacy-protected analyses. Certainly, the instability source of the studied $h_{vdc}(s)$ tuning cases is positioned at RE with the DC control loop to be improved in priority to avoid the unstable mode. The power control can also be selected to tune for $h_{vdc}(s)=30+1/0.05s$ when transformers are configured since it acts on the coupling between AC and DC terminals.

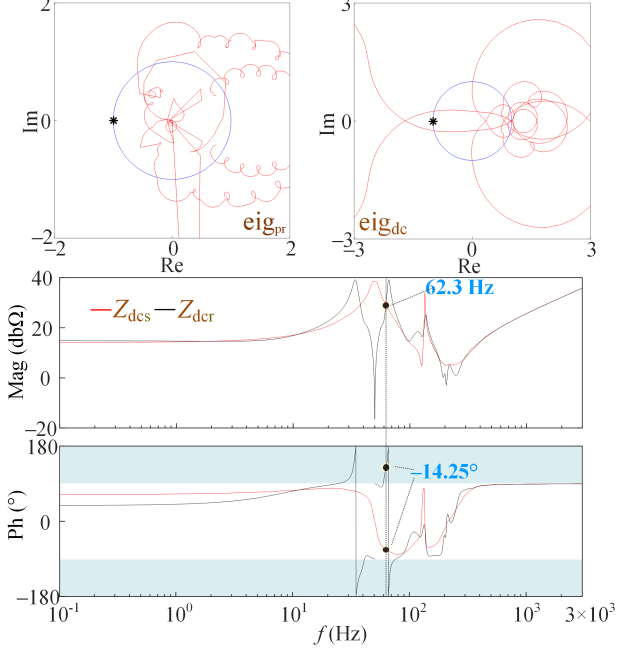

Fig. 16. Nyquist and Bode Plots ($h_{vdc}(s)=0.01+1/0.005s$).

### C. Case C: Step of P

Finally, as a practical concern, the stability of $P$ steps from 0.5 pu to 1 pu, is focused on, where $L_{gs}^*$ is changed to 10 mH to emulate a weak grid. A group of heuristic simulations is given for each subsystem, which has the time logic as follows:

1) At $t=3$ s, $L_g$ remains at $L_g^*$, $k_{pP}$ is set to 2, and $P^*$ is set to 0.5 pu; 2) At $t=4$ s, $k_{pP}$ becomes 2.5; 3) At $t=5$ s, $P^*$ steps to 1 pu; 4) At $t=7$ s, $L_g$ becomes 0; 5) At $t=8$ s, $k_{pP}$ becomes 2; 6) At $t=9$ s, $L_g$ turns to $L_g^*$. Fig. 17 claims that $k_{pP}=2$ or 2.5 will make the system unstable when $P^*=1$ pu on the premise that only subsystem ② cannot stably operate, where the frequency of sustained oscillation increases with a larger $k_{pP}$ ($|f_p-f_1|=230$ Hz and 236 Hz when $k_{pP}=2$ and 2.5).

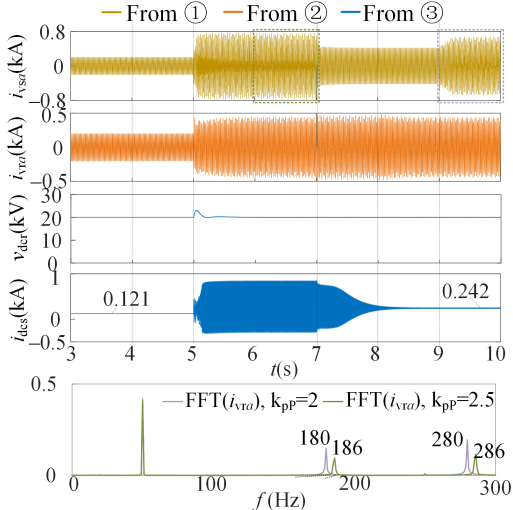

Fig. 17. Time domain simulation with the FFTs (case c).

Nevertheless, the frequency domain analyses using Fig. 18 and 19 illustrate that the same unstable mode can be identified at both DC and AC SE terminals when $k_{pP}=2$ but can only be identified at the AC SE terminal when $k_{pP}=2.5$, while the instabilities cannot be determined at the AC RE terminal (the simulation is neglected). The model frequency increases from 296 Hz to 299 Hz while PM decreases from −4.9° to −21.7°, which satisfies Fig. 17.

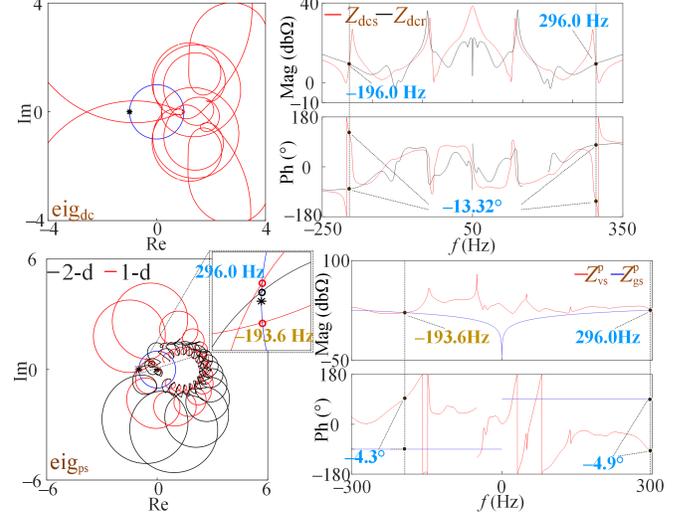

Fig. 18. Nyquist and Bode plots ($k_{pP}=2$).

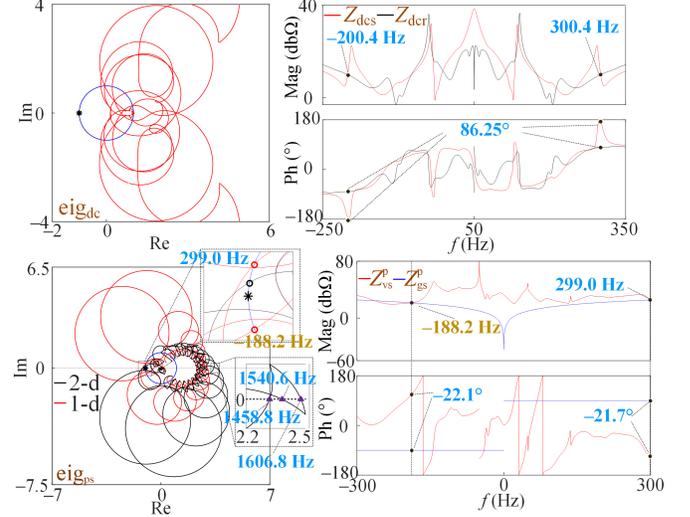

Fig. 19. Nyquist and Bode plots ($k_{pP}=2.5$).

Case c verifies the previous finding that supposing a specified terminal as the instability source for analysis may leave out the authentic unstable mode. Regarding the adaptivity of the simplified AC stability criteria for MMC from TL-VSC, more interesting findings are noted. First, the extra unstable modes appear in the 1-d PS loop criterion [26] as the red solid lines in the zoom-in of Fig. 18 and 19. Since $|f_p-f_1|=243.6$ Hz or 238.2 Hz deviates from the real 246 Hz or 249 Hz in two cases, it may affect the estimation of sustained oscillation frequency when $k_{pP}$ increases. In fact, 1-d and 2-d Nyquist criteria are only equivalent in marginal stability judging [26]. Furthermore, a practical determinant(Det($s$))-based criterion [27] is examined since the zeros and poles are difficult to obtain for IMs, as previously mentioned. By extracting a pair of conjugate right-

half plane poles from Det($s$), a *stable* mode satisfies:

$0.5\{\text{Im}[\text{Det}(s_{zc} + j2\pi)] - \text{Im}[\text{Det}(s_{zc} - j2\pi)]\}\cdot\text{Re}[\text{Det}(s_{zc})] > 0$ (24)

In (24), Im(Det($s$)) and Re(Det($s$)) separate the imaginary and real parts of Det($s$), and $s_{zc}$ is the zero-crossing angular frequency of Im(Det($s$)). In Fig. 20 when $k_{pP}$=2.5, the 1540.6 Hz mode is misjudged to be instability except for the expected 299 Hz, which is induced by incorrectly handling the time delay, as shown by the purple triangles shown in Fig. 12 (e). Overall, the 2-d Nyquist criterion is essential for AC stability analyses of MMC-BTB systems.

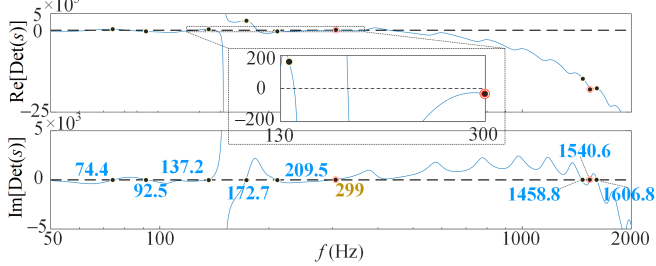

Fig. 20. Validation of the practical criterion [27] in the MMC-BTB systems considering the nonignorable time delay.

## VII. CONCLUSION

By focusing on an MMC-BTB system, this paper discusses several issues in extending the impedance-based analysis from a single MMC to multiterminal systems. With theoretical analyses and validation, the following conclusions are drawn:

1) A numerical method is proposed to rapidly and accurately obtain steady-state of the system regardless of whether transformers are configured, which helps unify the complete small-signal stability analysis in the frequency domain.

2) By clarifying the modeling idea, a generalized framework is offered to uniformly and succinctly develop AC/DC TIMs, which considers the typical control loops and the pure time delay in particular, and is verified by frequency scans.

3) Sensitivity analyses are adopted to emphasize the essential effects of the non-ideal transformer and ZSCC on the impedance characteristics, while the rationality of a novel simplified AIM considering the NS current control is comparatively studied.

4) Multiple case studies collectively confirm the necessity of executing AC/DC stability analyses with detailed IMs. The unstable mode can be identified at the AC or DC terminal of one MMC and the harmonics spread to the entire system, and it may not be suitable to replace Nyquist criteria with the simplified criteria in wideband stability analyses from TL-VSC to MMC.

Deeply understanding the modeling idea of this paper opens access to more complicated device-level and system-level impedance-based modeling and analysis. Future studies will focus on the couplings between multiple physical paths and the stability identification with multiple unstable subsystems in the multiterminal system.

## APPENDIX
## DETAILS OF CONTROL STAGE

The basic $\boldsymbol{J}$ can be decomposed according to Fig. 3, The elements not mentioned default 0 and the same below:

PS current control: $\boldsymbol{J}_1^p(s)(\text{x},\text{x}) = -0.5(1+g)\cdot[h_o\cdot h_d](s)$
$\cdot[h_i(s+j(x+g)\omega_l) + jgk_{de}]\cdot 2k_m k_i, \quad \text{x} = n \pm 1,$

NS current control: $\boldsymbol{J}_1^n(s)(\text{x},\text{x}) = -0.5(1+g)\cdot[h_o\cdot h_d](s+j2g\omega_l)$
$\cdot[h_i(s+j(x-g)\omega_l) - jgk_{de}]\cdot 2k_m k_i, \quad \text{x} = n \pm 1,$ (25)

CCSC: $\boldsymbol{J}_2^n(s)(\text{x},\text{x}) = -0.5(1-g)\cdot h_d(s)$
$\cdot[h_{CCSC}(s+j(x-2g)\omega_l) - j2gk_{de}]\cdot k_m k_i, \quad \text{x} = n \pm 2,$

$h_o(s-j\omega_l) = 0.5(1+e^{-sT_l/4}), \; h_d(s-j\omega_l) = e^{-sT_d}.$

The feedforward part of $\boldsymbol{K}$ is formed by two submatrices:

PS/NS current control: $\boldsymbol{K}_{1f}^p(s)/\boldsymbol{K}_{1f}^n(s)(\text{x},\text{x}) = k_m k_v k_f$
$\cdot e^{-\text{ang}(\boldsymbol{v}_v\langle 1\rangle)}\cdot\{[h_o\cdot h_d](s)/[h_o\cdot h_d](s+j2g\omega_l)\}, \quad \text{x} = n \pm 1.$ (26)

The influences on $\Delta \boldsymbol{m}$ contributed by PLL are divided into:

PSCC: $\boldsymbol{K}_{1PLL}^p(\text{x},\text{y})(s) = w\cdot[h_{PLL}^i(s) + h_{PLL}^m(s) + h_{PLL}^{v_v}(s)]$
$\cdot h_{PLL}(s), \quad \text{x} = n \pm 1, \; \text{y} = n \pm 1,$

NSCC: $\boldsymbol{K}_{1PLL}^n(\text{x},\text{y})(s) = w\cdot[h_{PLL}^i(s) + h_{PLL}^{v_v}(s)]$
$\cdot h_{PLL}(s), \quad \text{x} = n \pm 1, \; \text{y} = n \pm 1,$

CCSC: $\boldsymbol{K}_{2PLL}^n(\text{x},\text{y})(s) = w\cdot[h_{PLL}^m(s)]$
$\cdot h_{PLL}(s), \quad \text{x} = n \pm 2, \; \text{y} = n \pm 1,$ (27)

$w = 0.5\cdot(\text{x}-n)\cdot l\cdot\text{sign}[(\text{x}-n)\cdot(\text{y}-n)], \; h_{PLL}^m(s) = -2\boldsymbol{m}\langle|\text{x}-n|\rangle,$

$h_{PLL}^{v_v}(s+jg(\text{x}-n)\omega_l) = -k_m k_f k_v \cdot 2\boldsymbol{v}_v\langle|\text{x}-n|\rangle\cdot[h_o\cdot h_d](s),$

$h_{PLL}^i(s+jg(\text{x}-n)\omega_l) = 2k_m k_i \cdot 2\boldsymbol{i}\langle|\text{x}-n|\rangle$
$\cdot[h_o\cdot h_d](s)\cdot[h_i(s) - jg\cdot(\text{x}-n)k_{de}].$

$\Delta \boldsymbol{d}$ and $\Delta \boldsymbol{q}$ are different for each end as (x=$n$, y=$n\pm 1$):

$\Delta \boldsymbol{d}_r = \text{diag}[h_d(s+jg(\text{x}-n)\omega_l)]\cdot \boldsymbol{U}\cdot\Delta \boldsymbol{v}_{dc},$
$\Delta \boldsymbol{q}_r = \text{diag}[h_d(s+jg(\text{x}-n)\omega_l)]\cdot \boldsymbol{R}\cdot\Delta \boldsymbol{v}_l,$
$\Delta \boldsymbol{d}_s = \text{diag}\{[h_o\cdot h_d](s+jg(\text{x}-n)\omega_l)\}\cdot(\boldsymbol{S}\Delta \boldsymbol{v}_l + \boldsymbol{T}\Delta \boldsymbol{i}),$
$\Delta \boldsymbol{q}_s = \text{diag}\{[h_o\cdot h_d](s+jg(\text{x}-n)\omega_l)\}\cdot(\boldsymbol{W}\Delta \boldsymbol{v}_l + \boldsymbol{V}\Delta \boldsymbol{i}),$ (28)

$\boldsymbol{W}(\text{x},\text{y}) = j6g\boldsymbol{i}\langle-(\text{y}-n)\rangle/k_a, \; \boldsymbol{V}(\text{x},\text{y}) = -j3g\boldsymbol{v}_v\langle-(\text{y}-n)\rangle,$
$\boldsymbol{S}(\text{x},\text{y}) = -6\boldsymbol{i}\langle-(\text{y}-n)\rangle/k_a, \; \boldsymbol{T}(\text{x},\text{y}) = -3\boldsymbol{v}_v\langle-(\text{y}-n)\rangle,$
$\boldsymbol{R}(\text{x},\text{y}) = 1.5[e^{-(s-j\omega_l)T_l} - 1]\cdot\boldsymbol{v}_v\langle-(\text{y}-n)\rangle/[V_b\cdot(s-j\omega_l)T_l].$

$\boldsymbol{E}$ and $\boldsymbol{F}$ are given for each end as (x=$n$):

$\boldsymbol{E}_r(s)(\text{x},\text{x}) = k_{v_{dc}}k_m\cdot[h_{v_{dc}}\cdot h_i](s-j\omega_l),$
$\boldsymbol{E}_s(s)(\text{x},\text{x}) = k_{PQ}k_m\cdot[h_{PQ}\cdot h_i](s-j\omega_l),$ (29)
$\boldsymbol{F}_r(s)(\text{x},\text{x}) = -jg k_{v_{RMS}}k_m\cdot[h_{v_{RMS}}\cdot h_i](s-j\omega_l),$
$\boldsymbol{F}_s(s)(\text{x},\text{x}) = -jg k_{PQ}k_m\cdot[h_{PQ}\cdot h_i](s-j\omega_l).$

$\boldsymbol{B}_i, \boldsymbol{B}_{v_{dc}}$, and $\boldsymbol{B}_{v_v}$ are expressed for each end as:

$\boldsymbol{B}_{v_{dc}r} = -\boldsymbol{E}_r'', \; \boldsymbol{B}_{ir} = -\boldsymbol{J}, \; \boldsymbol{B}_{v_v r} = -(\boldsymbol{K} + \boldsymbol{F}_r''\boldsymbol{R}), \; \boldsymbol{B}_{v_{dc}s} = \boldsymbol{O},$
$\boldsymbol{B}_{is} = -[\boldsymbol{J} + (\boldsymbol{F}_s'\boldsymbol{V} + \boldsymbol{E}_s'\boldsymbol{T})/k_a], \; \boldsymbol{B}_{v_v s} = -(\boldsymbol{K} + \boldsymbol{F}_s'\boldsymbol{W} + \boldsymbol{E}_s'\boldsymbol{S}).$ (30)

where the double (single) apostrophe means to integrate the matrices with $h_d(s)$ ($[h_o\cdot h_d](s)$) of $\Delta \boldsymbol{d}$ or $\Delta \boldsymbol{q}$ into $\boldsymbol{E}$ or $\boldsymbol{F}$.

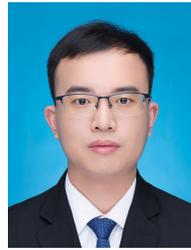

**Chongbin Zhao** (Student Member, IEEE) received the B.S. degree in electrical engineering from Tsinghua University, Beijing, China, in 2019, where he is currently working towards the Ph.D. degree. He is also a visiting scholar at Rensselaer Polytechnic Institute, Troy, NY, United States, from January 2023. His research interests include power quality analysis and control, and emerging converter-driven power system stability analysis and control.

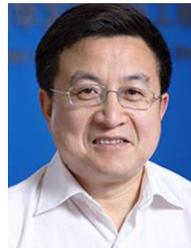

**Qirong Jiang** received the B.S. and Ph.D. degrees in electrical engineering from Tsinghua University, Beijing, China, in 1992 and 1997, respectively. In 1997, he was a Lecturer with the Department of Electrical Engineering, Tsinghua University, where he later became an Associate Professor in 1999. Since 2006, he has been a Professor. His research interests include power system analysis and control, modeling and control of flexible ac transmission systems, power-quality analysis and mitigation, power-electronic equipment, and renewable energy power conversion.